# CLASSIFY PARTICIPANTS IN ONLINE COMMUNITIES

Xuequn Wang[1] and Yanjun Yu[2]

[1]Department of Management, Information Systems & Entrepreneurship, Washington State University, Pullman, WA, USA
xuequnwang@wsu.edu
[2]Department of Management Information Systems, Southern University at New Orleans, New Orleans, LA, USA
yyu@suno.edu

## ABSTRACT

*As online communities become increasingly popular, researchers have tried to examine participating activities in online communities as well as how to sustain online communities. However, relatively few studies have tried to understand what kinds of participants constitute online communities. In this study, we try to contribute online community research by developing "common language" to classify different participants in online communities. Specifically, we argue that the previous way to classify participants is not sufficient and accurate, and we propose a continuum to classify participants based on participants' overall trend of posting activities. In order to further online community research, we also propose potential directions for future studies.*

## 1. INTRODUCTION

Advances in information and communication technologies (ICT) are changing individuals' life, and a new trend since 21st century is online communities. Online communities enable individuals not only to post user generated content (UGC) such as feedback, opinions and personal information [1], but also exchange information with others who they may never meet. Researchers have examined how online communities shape individuals' daily life and what the implications are for practitioners and researchers [19]; how to transfer users of collaborative websites into contributors [1]; and the needs, i.e., recognition and appreciation of their contribution, promotional opportunities or because of sense of responsibility, of the knowledge sharing in online communities inside or outside of an organization [18].

While previous studies have tried to understand participating activities in online communities as well as how to sustain online communities [11, 32], relatively few studies have tried to classify participants in online communities [10, 28]. Further, previous studies mainly focus on those participants who answer questions or contribute content to online communities, there are no established terminologies for participants who engage in other activities (e.g., make comment, provide feedback). Although there is a kind of individuals called lurker, the term "lurker" may not be suitable in many contexts; given that those individuals do engage in certain activities in online communities, and these activities can also be important to online communities.

Therefore, in this paper, we try to contribute online community research by classifying different participants in online communities. In our study we use individuals and participants interchangeably, and use information to refer to useful and valuable content, and content to refer to all of the posts from online communities. There are different types of online communities [7], and individuals go to online communities for different purposes. Our study only focuses on those online communities where participants share and exchange information with one another [14], and other types of online communities (e.g., discussion board for online education, online social networks) are beyond the scope of this study. We admit that our study is only an initial step toward deeper understanding of participants in online communities.

The rest of the paper is organized as follows. First, we conduct a brief review of previous literature. We then analyze posts from one online community, and classify them into different types. Next, we try to clarify the exact meanings of contribute and lurk. Based on these discussions, we propose a continuum to classify participants in online communities. Finally, we discuss the implications for future studies.

## 2. AN OVERVIEW OF PARTICIPANTS IN ONLINE COMMUNITIES

### 2.1 Contributors

Many previous studies classify (or just mentions) two types of participants in online communities: contributors (or poster) and lurkers [22, 28]. While the concept of contributors receives little debate, there are competing ideas about who the lurkers really are. For example, Tedjamulia et al. [28] refer lurkers to two types of individuals: (1) those who just view posts; (2) those who do not find a specific type of information and ask for it. In another study, Takahashi et al. [27] refer lurkers to those who do not post any message. Preece et al. [22, p.202] even go one step further and argue that a lurker is "someone who has never posted in the community to which he/she belongs".

Therefore, there are debates about the definition of lurker. To better understand the concept of contributor and lurker, we first go through *Merriam-Webster dictionary* to find out what contribute and lurk exactly mean. There are three meanings for contribute: 1) to give or supply in common with others (e.g., contribute money to a cause); 2) to supply (as an article) for a publication; 3) to play a significant part in bringing about an end or result (e.g., many players have contributed to the team's success).

Based on abovementioned meanings, the meaning of contributor in online communities is derived from the second and third meaning. Specifically, previous literature considers participants providing new information play a significant part in sustaining online communities, and therefore calls them "contributors". After all, without valuable information, online communities cannot attract other participants and are likely to fail [32].

However, these discussions bring another question: are there other ways to help sustain online communities by providing some "content"? Online communities allow individuals to interact with each other [2]. In such a conversational environment, individuals "create and share knowledge through dialog with questions and answers" [30, p. 266]. Phang et al. [20, p. 721] also argue that the value of online communities depends on "ongoing participation in terms of two key activities i.e., knowledge seeking and contribution". Therefore, participants who ask questions may also help sustain online community by finding what information is not available and initiating interaction among participants. In other words, besides contributors, there are other types of participants who can also help sustain online communities.

In fact, recent studies begin to pay more attention to other types of participants. For example, Majchrzak et al. [16] categorize contributors into two kinds: "synthesizers" and "adders". Synthesizers are those mainly to influence other (wiki) users (e.g., influence the task by finding a new solution), while adders mainly provide new information. Clearly, synthesizers also help sustain online communities. Besides, information that is useful to one individual may be distracting to another [3], and other types of participants can help clarify the usefulness of the information.

To summarize, other participants actively engaging in online communities can also supply some "content" and help sustain online communities. Therefore, previous definitions on contributor [28] are limited, and participants in online communities need to be classified further.

## 2.1 Lurkers

Next we examine the definition of lurker. According to *Merriam-Webster* dictionary, there are four meanings for lurk: 1) to lie in wait in a place of concealment; 2) to move furtively or inconspicuously; 3) to persist in staying; 4) to read messages on an Internet discussion forum (as a newsgroup or chat room) without contributing.

Based on these meanings mentioned above, lurkers are those who just read posts in online communities (fourth meaning). That meaning is derived from the first meaning of lurk, stating that lurkers are those who hide from other individuals in online communities. This explanation is consistent with the first kind of lurkers from Tedjamulia et al. [28], but not consistent with their second one.

Researchers also realize that there are different types of lurkers. For example, Dennen [6] argues that there are four different types of lurkers: 1) individuals who do not post anything; 2) individuals who read but do not post in a given discussion; 3) individuals who are new and not ready to post; 4) individuals who never want to post. The result of this study implies that individuals go to online communities with different purposes, and the overall trend of activities from individuals are probably more accurate to classify their roles in online communities since the overall trend of activities can better represent individuals' purposes to go to online communities. In other words, participants who are new to the online communities should be treated as lurkers in a traditional sense because these participants are not ready yet. Once these participants are familiar with the online community, they are likely to contribute and post useful information.

To summarize, the definition of contributors and lurkers remains debatable, and we need to classify participants in more depth. To achieve that goal, we first try to understand the activities in online communities by examining posts in an online community and develop a schema to classify these posts.

## 3. EXAMINE ACTIVITIES IN ONLINE COMMUNITIES

In this part, we examine activities (posts) in online communities. Although some studies consider electronic environment suffering from social cue deficiencies [23, 25, 26], other studies also find that even "plain" media such as email can contain rich communication information [17]. Therefore, it is fairly important to understand how individuals participants in online communities.

In a recent study, Hrastinski [10] lists six ways of participants: 1) participation as accessing environments; 2) participation as writing; 3) participation as quality writing; 4) participation as writing and reading; 5) participation as actual and perceived writing; 6) participation as taking part and joining in a dialogue. Based on this list, writing and posting is an important way to participate, and we try to classify different posts from participants.

To examine activities and classify posts, we select DIS Discussion Forums (www.disboards.com) as the sources of posts. DIS Discussion Forum (DIS) is one of the largest and liveliest Disney forums. Participants in DIS share information and discuss a variety of topics such as Disney theme parks, resorts and Disney Cruise Line with each other. Therefore, DIS is consistent with our focus and is a good example to examine online communities where participants share information.

We select post as the analysis unit since the basic unit in online communities is a "comment-by-participant-on-topic-at-time" [31]. Also, we focus on dialog in this study, which is one of the two paradigms for the sharing and distribution of information in online communities [31]. By analyzing the posts from DIS during November and December in 2010, we classify posts of DSI into seven categories (see Table 1):

Table 1. Post Types

| Post Type | Short description | Position in a thread | Affect | Explicit or Implicit |
|---|---|---|---|---|
| Post Type I | Posts which provide new information | In the beginning | Neutral | Explicit |
| Post Type II | Posts which ask questions | Anywhere | Neutral | Explicit |
| Post Type III | Response posts which answer questions | In the middle | Neutral | Explicit |
| Post Type IV | Response posts which provide feedback | In the middle | Neutral | Explicit |
| Post Type V | Response posts which thank for help | In the middle | Positive | Explicit |
| Post Type VI | Response posts which say something bad | In the middle | Negative | Explicit |
| Post Type N | Viewing without posting | N/A | Negative | Implicit |

### 3.1 Posts which provide new information (Post Type I):

When individuals have some new information to share, they may put the information in a single post. The post usually appears in the beginning of a thread. We refer this kind of posts to "Post Type I". "Post Type I" provides interesting or valuable information and is usually relevant to other participants from the same online community. For example, here is an example from DIS:

"*This thread is intended to discuss most aspects of Walt Disney World (WDW) ticketing…*"

In fact, that post is so useful that the administrator put it at the top of one subforum.

### 3.2 Posts which ask questions (Post Type II):

Online communities cannot contain every piece of information for which their participants look. Even if the relevant information does exist, that piece of information may be quite difficult for participants to locate. Therefore, individuals sometime (start threads and) ask questions. We refer this kind of posts to "Post Type II". For example, here is a post from DIS:

"*Hi! I'm planning my first trip to Disney with my family. I would like to be able to find out which characters will be where and when, so I can make the most of each park we visit. I'm sure in the MK they will be everywhere, but I would especially love to know when Mike and Sully and the Cars characters will be appearing in HS.*"

Since participants from the same online community usually share similar interests, other participants may also want to ask the same or similar question. If the information required is indeed not available, others may answer the question and thus increase the information base of online communities (refer to Post Type III).

### 3.3 Response posts which answer questions (Post Type III):

When someone asks a question, other participants may know the answer and response to that question, by either providing relevant information directly or referring to previous posts if the answer already exists. We refer this kind of posts to "Post Type III". For example, here is one of answers from DIS to the example question of the Post Type II:

"*You can also ask at guest services once you are in the park. They can tell you when where and if a character will be there that day.*"

Post Type III is often context-based. However, Post Type III does contain valuable and useful information, and increases the information base of online communities, which you can see from the above example. Post Type III also functions as the intermediary connecting participants who have questions to those who know answers.

### 3.4 Response posts which provide feedback (Post Type IV):

After participants who ask questions get answers, they may not feel that their questions are really answered. The reason may be that individuals who give answers misunderstand the questions, or that those participants who ask the questions need clarification. In these situations, participants who ask questions provide feedback through response posts. We refer this kind of posts to "Post Type IV". "Post Type IV" can also come from individuals who try to answer questions, but find that they need more details. For example, here are three sequential posts from DIS.

A: "*…can anyone tell me what time the Jedi Academy is held each day??*" (Post Type II)

B: "*My strategy would be to arrive at Rope Drop, ride TSM or fastpass then sprint to Jedi Academy and wait.*" (Post Type III)

A: "*Where exactly is this in the park? My 7y/o is SW obsessed and has every light saber known to man, lol*" (Post Type IV)

Although Post Type IV can also be a question, it should not be equal to Post Type II. The purpose of Post Type II is to get some new information, while that of Post Type IV is to clarify and achieve mutual understanding.

Another possibility is that someone posts something and others express what they think about it. For example:

A: "*Check out this website its like people of walmart only at Disneyworld, to funny, http://www.peopleofthepark.com/*" (Post Type I)

B: "*People of the Park is interesting, but a little on the mean side. Maybe I'm just scared I'll show up on there!*" (Post Type IV)

C: "*Me, too. I can think of one specific instance this past July. It has to do with a matching shirt I was guilted into wearing.*" (Post Type IV)

Post Type IV is usually the middle stage between Post Type II and Post Type III. Sometimes Post Type IV is not needed, while sometimes many Post Type IV posts are needed to clarify the contexts. These posts may not contain any valuable information, but are important to make "conversation" between participants keep going and get answers finally. Post Types I to IV are affective neural, from the perspective of contributors[1].

### 3.5 Response posts which thank for help (Post Type V):

When participants who ask questions receive answers that they want, they probably appreciate the help and effort from others. We refer this kind of posts to "Post Type V". For example, here is an example from DIS:

"*awesome! thanks for the help. I haven't checked the links yet, but heading there now. Great maps, Robo!*"

Based on the above example, we can tell that the Post Type V post usually contains positive affect to contributors.

### 3.6 Response posts which say something bad (Post Type VI):

Sometimes participants may feel angry (about previous posts) and throw bad words into online communities. We refer this kind of posts to "Post Type VI". Post Type VI post usually contains negative affect to contributors. Although we did not find such type of posts in DIS during analyzing, we do come across them occasionally from other contexts and decide to include this type to make our classification more complete.

### 3.7 Viewing without posting (Post Type N)

Individuals may simply view rather than post something in the online communities. We refer this kind of "posts" to "Post Type N". Post Type N is not a real post. However, some changes may occur after viewing. For example, some online communities have the function to count how many times individuals have viewed a thread. So even if individuals just view posts, this kind of "post" will increase the count of viewing by one. Some contributors may feel honoured that others take time to go through their posts even without responses. However, we consider Post Type N to be implicitly affectively negative to contributors for the following reasons: in many cases administrators encourage responses, and many participants who contributed information in the online communities ask others how they feel about the information were posted. If other participants just view the posts without providing any response or feedback, the online community contributors don't know whether the information they provide need further modification or clarification. Thus, viewing without posting do not help contributors understand how useful their information is.

We admit that just examining one online community is a limitation of this study. Nevertheless, we believe that most of the post types identified can be also applied to other contexts. Please also note that some posts may be multi-type. For example, one post can both comment on the previous posts and raise a new question.

---

[1] Given that content contribution is one of the central interests in online community research [32], we choose to view posts from the perspective of contributors.

## 4. A CONTINUUM OF PARTICIPANTS IN ONLINE COMMUNITIES

Takahashi et al. [27] argue that there are five factors which can be used to classify participants from online communities, including expectations of purpose of an online community, stance on participation, personal specialty and interests, attitude towards information handling, and awareness of the existence of others. To separate individuals into meaningful subdivision, as well as keep the classification parsimonious enough, we argue that classification of participants in online communities should be based on their overall activities (or tendency) in a certain time period, and the value[2] of different types of participants to a certain online community are indeed along a continuum (refer to figure 1). This criterion is consistent with expectations of purpose and stance on participation from Takahashi et al. [27]. Thus, we classify all individuals from a certain online community (we refer it to X Community in this part) into seven categories.

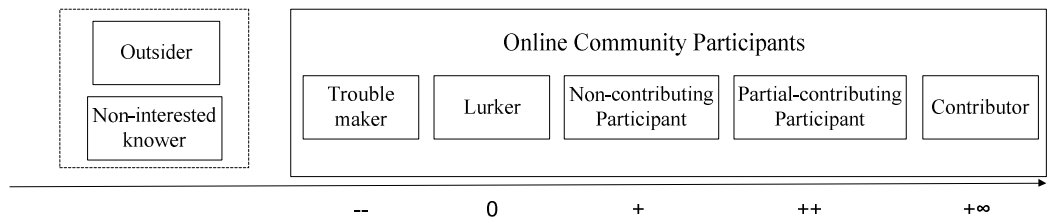

A Continuum of Online Community Participants

Figure 1. A Continuum of Participants in Online Communities

***Outsider***: Individuals who do not know X community. They may or may not be interested in the content from the X community.

***Non-interested knower***: Individuals who know X community but never browse its posts. They are currently not interested in the content from the X community. They may become interested in the content from the X community in the future. This concept is consistent with the concept of shirkers from Egan et al. [8].

***Trouble maker***: Individuals who post bad words (Post Type VI) in X community. Once individuals put Post Type VI posts, which can hurt the X community by letting other participants feel bad. Therefore, even if these individuals may provide new information sometimes (Post Type I or III), they are not desired by the X community because other participants may leave the X community because of Post Type VI posts.

***Lurker***: Individuals who just consume without providing any new information. In other words, they just "put" Post Type N. While they increase the total number of visiting of X community, they also generate Post Type N posts. Therefore, their overall value to the X community is roughly zero.

---

[2] We define value as the degree to which individuals help sustain online communities.

***Non-contributing Participants***: Individuals who do not provide any new information, but consume information or participate in other activities. They can ask questions (Post Type II), give feedback (Post Type IV), and thank others (Post Type V).

***Partial-contributing Participant***: Individuals who contribute sometimes, but they usually consume information from X community or participate in other activities. These participants still provide new information (Post Type I) or answer questions (Post Type III) sometimes, but they usually read posts or participate in other activities (Post Type II, IV, and V).

***Contributor***: Individuals who provide new information (Post Type I and III) regularly. They do spend some time browsing posts, but their main activities are contributing.

In the continuum, we try to avoid the confusion of the "contribute" concept discussed early by explicitly adding the term "participate". Thus, we differentiate other activities (Post Type II, IV, and V) from providing new information (Post Type I and III), and narrow the meaning of "contribute". We put non-interested knower closer to the X community than outsider because the non-interested knowers at least know the existence of the X community.

In a certain study researchers can define the exact meaning of time period according to their research questions and interests. For example, if one study wants to examine participants' monthly activity pattern, then the time period is one month and classification of participants can accord to their overall activities in every month.

Although it is tempting to have individuals who just contribute, this type of individuals cannot exist in the reality. To answer questions, individuals have to go through other participants' posts. Even for those who often provide information spontaneously (Post Type I), these contributors also want to read others' responses to see how others think about their posts or if there is something missing in their posts.

## 5. WHAT ONLINE COMMUNITIES NEED IN DIFFERENT STAGES

Given that only partial-contributing participants and contributors actively contribute in online communities, these two kinds of participants are probably the most desired in online communities[3]. However, online communities may vary in the stage of their life cycles [21]. Therefore, online communities may want to put different emphasis on different types of participants in different stages. Table 2 lists the short description of each stage, and Figure 2 shows how the number of participants changes in each stage. Below we provide a more detailed discussion of each stage.

---

[3] Here we do not mean that non-contributing participants are not desired at all. As we discussed early, these participants do have values to the online community, but their value is relatively less than that of partial-contributing participants and contributors.

Table 2. Different Stages Of Online Community Life Cycle

| Stage | Short description |
| --- | --- |
| Attraction | Online communities start to launch; Contributors provide much interesting information, and individuals from outside are attracted. |
| Build-Up | As more and more individuals are attracted, they begin to build up the relationship with online communities and participate regularly. |
| Maintenance | The number of participants passes a critical mass; the number of daily posts is so large that online communities focus more on the maintenance of posts. |
| Deterioration/ End | The rate of providing new information is decreasing; participants start to leave online communities and stop participating. |

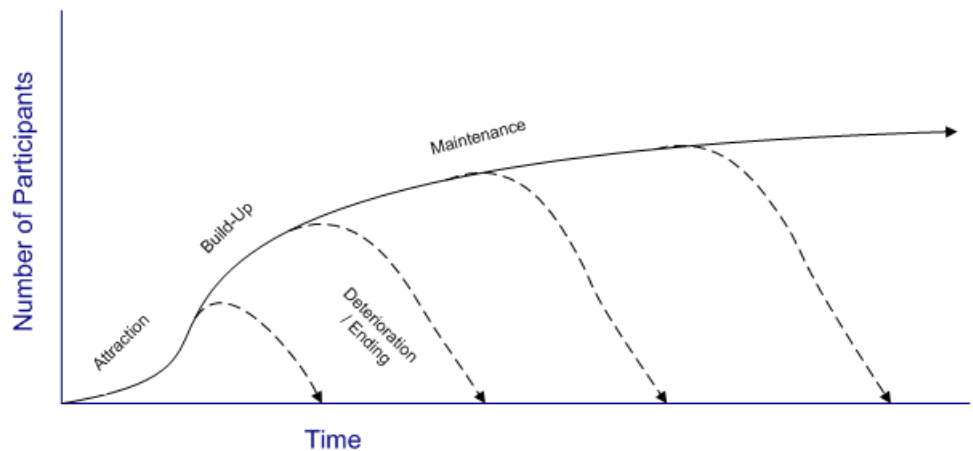

Figure 2.  Online Communities in Different Stages

During the early stages (Attraction and Build-Up), online communities just start and need to attract more individuals. Since the useful and interesting information of the online communities is the key to attract individuals outside, contributors are most desired [32]. During these stages, the number of participants is not large enough and there is not enough information in the online communities. If most participants just comment instead of providing new information, the information from the online communities may become out of date soon and participants may find that there is not much to gain [22]. Therefore, partial-contributing participants are less desired during this stage, and the online communities need many contributors who contribute a lot and regularly.

When online communities become mature, and the size of participants becomes large enough (exceeds a critical mass), the emphasis of the online communities probably switches to maintenance, and other types of participants are more desired than those of early stages. Butler [3] points out that the main focus of online communities comes to the maintenance with the growth of community size. In Wiki context, Kittur et al. [13] find that as the size of Wikipedia grows, the rate of creating new information is decreasing, while the amount of maintenance work is increasing. Viégas et al. [29] also notice that the fastest growing area of Wikipedia is group coordination, which Wikipedia community places a strong emphasis on, and group coordination supports strategic planning of edits and enforcement of standard guideline. Therefore, in the stage of maintenance, non-contributing and partial contributing participants become more important.

International Journal of Managing Information Technology (IJMIT) Vol.4, No.1, February 2012

They may provide little new information, but they actively engage in maintenance-related activities.

Non-contributing and partial contributing participants can also be important to contributors. Contributors need other individuals to participate (e.g., join the "conversation") to help them keep contributing. Butler et al. [4] point out that the most fundamental investment in online communities is active participation, including posting content, responding to posts, and organizing discussion. From a collaboration standpoint [5], contributors need confirmation from other individuals once they provide some information. From a resource standpoint, membership size is a kind of resource: audience resource. When contributors provide new information, they also look for audiences' ideas. In other words, they compete with audience resource [3], and need other individuals to recognize their effort of providing new information [11, 12, 32].

Further, Non-contributing and partial contributing participants can help identify what pieces of new information are needed. Hahn and Subramani [9] point out that knowing a prior what information will be requested is quite challenging. That is especially true for online communities. In such a context, Post Type II can be useful. These arguments are consistent with the concept of conversational knowledge management [30]. Wagner and Bolloju [31] also argue that online communities carry out knowledge creation and sharing through a process of discussion with questions and answers (discussion forum), collaborative editing and maintenance (Wikipedia), or storytelling (weblogs).

Therefore, the situation becomes much more complex during the maintenance stage. While online communities still need Post I and III posts, other types of post (Post Type II, IV and V) are more desired than Post I and III in later stages, and the online communities need to balance these activities carefully. Meanwhile, post type VI is not desired and even prohibited. Here, editorial content control can screen out these unwelcomed activities [3]. Therefore, even after the online communities move to the stage of maintenance, these communities are not guaranteed to success forever. Without careful monitoring and maintenance, online communities may deteriorate and end.

Although lurkers are not desired, we do not mean that individuals should not view and consume information. In fact, consuming information is one type of participations, and online communities will not remain viable if individuals do not regularly consume the information that contributors provide [4]. However, after individuals consume information, they are desired to give contributors feedback (Post Type IV) or thank contributors (Post Type V). Viewing with posting (Post Type N) may not be desired.

The discussion above focuses on those participants who are active in online communities in a certain way. Does it mean that there is nothing to do with outsiders, non-interested knowers and lurkers? Probably not, and we argue that those three types of individuals can be as interesting as partial-contributing participants and contributors. In the next part we provide a few potential research directions for those types of individuals.

## 6. IMPLICATIONS FOR FUTURE STUDIES

In this part, we propose a few potential research directions, which we hope can give some insights to further online community research. Previous research mainly focuses on the factors to motivate individuals to contribute. In other words, researchers are mainly interested in "converting" lurkers, non-contributing participants to partial-contributing participants and even contributors. While we fully agree that this stream of research is quite important for online community research, we also realize that mainly focusing on contributors may be limited.

Before online communities even encourage individuals to actively participate, these individuals need to know the online communities. Therefore, one interesting question can be "What initiatives can help online communities be known by more individuals?" After all, potential contributors cannot become resource of online communities if they do not even know those online communities.

Many online communities are interest-based, and their participants generally share similar interests. Therefore, non-interested knowers are probably not interested in the content of that online community[4]. However, while some online communities organize all content under a shared theme (e.g., all subforums of DIS are related to Disney Park), others may have several loosely organized subforums and cover many different areas of content such as sports, movie, and politics (e.g., the discussion board in sina.com). In such a context, individuals actively participating in a few subforums may never visit other subforums. Therefore, one interesting question is "Is it possible (or beneficial) to include more subforums and attract more individuals?" Clearly, adding another subforum can potentially increase the size of participants. However, the cost of organizing and maintaining can also go up. Managers of online communities should carefully balance the cost and the benefit when considering such decisions.

Even if lurkers just consume information without posting, online communities can still impact lurkers by influencing their daily life. Therefore, researchers may want to examine lurkers' daily life outside online communities, and divide lurkers into different sub-types, based on the impact of online communities. For example, Willett [33] divides lurkers into two kinds: active lurkers who make direct contact with posters or propagate information gained, and passive lurkers who only read for their own use. Based on Willet's study, Takahashi et al. [27] further divide active lurkers into two types: active lurkers as propagators, who "propagates information or knowledge gained from an online community to others outside it", and active lurkers as practitioners who "uses such information or knowledge in their own or organizational activities". They also classify passive lurkers into two types: active lurker candidates and the persistent lurkers based on "whether the online community affects the lurker's thought". Therefore, one interesting question is "How can online communities impact lurkers' daily behaviours?"

Recent studies have begun to understand the effect of online communities on lurkers in the context of learning. For example, Dennen [6] showed that students learned through reading messages in the online discussion board. This research direction is consistent with knowledge management literature. For example Kankanhalli et al. [11] argue that successful knowledge management systems need knowledge seekers to use and reuse the codified knowledge. Lindgren and Stenmark [15] also argue that individuals can hint their friends about what they think interesting and useful by "word-of-mouth" [24]. Therefore, understanding the effect of online communities on lurkers is quite important.

## 7. CONCLUSIONS

Online community research becomes increasingly popular. In this study, we try to develop terminologies to classify participants in online communities. To achieve this goal, we analyze the posts from one online community, and then classify participants along a value continuum. We then propose potential directions for future studies. We hope our study can further the research on online communities.

## REFERENCES

[1]. Abrouk, L., Gross-Amblard, D. & Cullot, N. (2010) "Community Detection in the Collaborative Web", *International Journal of Managing Information Technology*, Vol. 2, No. 4, pp1-9.

[2]. Brown, J. S., & Duguid, P. (1991) "Organizational Learning and Communities of Practice," *Organization Science*, Vol. 2, No. 1, pp40-57.

---

[4] Here the assumption is that the online community is matured and its content coverage is relatively stable. When online communities just begin, their content coverage can change substantially and those online communities are possible to attract non-interested knowers.

**Authors**

Xuequn (Alex) Wang is a Ph.D. Candidate from the Department of Management, Information Systems, & Entrepreneurship at Washington State University. He received his M.S. in Management Information Systems from Oklahoma State University, and a B.E. from Civil Aviation University of China. His research interests include psychometrics, knowledge management, online communities, and idea generation. His research has been presented at *Hawaii International Conference on System Sciences (HICSS)*, *Americas Conference on Information Systems (AMCIS)*, *Western Decision Sciences Institute (WDSI)*, and *JAIS Theory Development Workshop*.

Yanjun Yu is an assistant professor in the Department of Management Information Systems (MIS) at Southern University at New Orleans. She received her Ph.D. in MIS from Washington State University, a M.S. in Management and Information Technology from California State University Monterey Bay, and a B.A in Business English from Zhengzhou University, China. Her research interests include cross-culture study, global software outsourcing, communication and mobile commerce, etc.. Her research papers have appeared on *Electronic Markets*, *The Data Base for Advances in Information Systems* and *Communications of the Association for Information Systems (CAIS)*, and have been presented on many conferences such as *International Academy of Business and Public Administration Disciplines (IABPAD) Conference, the Americas Conference on Information Systems (AMCIS)*, *the Conference on Information Science Technology and Management (CISTM)*, *Western Decision Sciences Institute (WDSI)*, and *Pre-International Conference on Information Systems (ICIS)*, *ISAP Workshop*.